\documentclass{nature}
\usepackage{hyperref}
\usepackage[pdftex]{graphicx,color}
\usepackage{amsmath, amssymb}
\usepackage{graphicx}
\usepackage[english]{babel}
\usepackage{subcaption}

\definecolor{vale}{rgb}{0,0.5, 1.}
\definecolor{davi}{rgb}{0.2,0.3, 1.}

\newcommand{\mscript}[1]{{\mbox{\scriptsize #1}}}
\newcommand{\mtiny}[1]{{\mbox{\tiny #1}}}
\DeclareMathOperator\erf{erf}
\def\d{{\rm d}}

\title{Absence of a fundamental acceleration scale in galaxies}

\author{Davi C. Rodrigues$^{1,2}$, Valerio Marra$^{1,2}$, Antonino del Popolo$^{3,4,5}$ \&  Zahra Davari$^{6}$}

\makeatletter
\let\saved@includegraphics\includegraphics
\AtBeginDocument{\let\includegraphics\saved@includegraphics}
\renewenvironment*{figure}{\@float{figure}}{\end@float}
\makeatother

\begin{document}

\maketitle

\begin{affiliations}
 \item {\footnotesize Center for Astrophysics and Cosmology, CCE,  Federal University of Esp\'irito Santo, 29075-910, Vit\'oria, ES, Brazil.}
 \item {\footnotesize  Department of Physics, CCE,  Federal University of Esp\'irito Santo, 29075-910, Vit\'oria, ES, Brazil.}
 \item {\footnotesize  Dipartamento di Fisica e Astronomia, Universit\`a di Catania, Viale Andrea Doria 6, 95125 Catania, Italy.}
 \item I{\footnotesize  nstitute of Modern Physics, Chinese Academy of Sciences, Post Office Box 31, Lanzhou 730000, People's Republic of China.}
 \item {\footnotesize  INFN sezione di Catania, Via S. Sofia 64, 95123 Catania, Italy.}
 \item {\footnotesize  Department of Physics, Bu Ali Sina University, Hamedan, Iran.}
\end{affiliations}


\begin{quotation}
{\it  \noindent This version does not match the published version at Nature Astronomy. \\ The published version can be freely found at \textcolor{davi}{\url{https://rdcu.be/ZXNT}}}.	
\end{quotation}

\bigskip

\begin{abstract}
The Radial Acceleration Relation\cite{McGaugh:2016leg} confirms that a nontrivial acceleration scale can be found in the average internal dynamics of galaxies.
The existence of such a scale is not obvious as far as the standard cosmological model is concerned, and it has been interpreted as a possible sign of modified gravity\cite{Lelli:2017vgz, Smolin:2017kkb}.
The implications could be profound: it could in principle explain galactic dynamics without  large amounts of  yet-undetected dark matter\cite{Akerib:2016vxi, Aprile:2017iyp} and  address issues that the standard cosmological model faces at galactic scales\cite{0521857937, DelPopolo:2013qba, DelPopolo:2016emo, Liu:2017drf, Rodrigues:2017vto}.
Here, we consider 193 disk galaxies from the  SPARC\cite{2016AJ....152..157L} and  THINGS\cite{2008AJ....136.2563W, 2008AJ....136.2648D} databases and, using Bayesian inference, we show that the probability of existence of a fundamental acceleration that is common to all the galaxies is essentially zero:
the $p$-value is smaller than $10^{-20}$ or, equivalently, the null hypothesis is rejected at more than 10$\sigma$.
We conclude that the acceleration scale unveiled by the Radial Acceleration Relation is of  emergent nature, possibly  caused by a complex interplay between baryons and dark matter. In particular, the MOND theory\cite{1983ApJ...270..371M, Sanders:2002pf, Famaey:2011kh, Milgrom:2015ema}, or any other theory that behaves like it at galactic scales, is ruled out as a fundamental theory for galaxies at more than 10$\sigma$.  
\end{abstract}

Dark matter is currently one of the main mysteries of the universe. There are many strong indirect evidences that support its existence, from galactic rotation curves and galaxy cluster dynamics to cosmological structure formation and cosmic microwave background anisotropies, but there is yet no sign of a direct detection\cite{Akerib:2016vxi, Aprile:2017iyp}.
Moreover, at the scales of galaxies, there is tension between the theoretically expected dark matter distribution in the universe (from the standard cosmological model, $\Lambda$CDM) and its indirectly observed distribution\cite{0521857937, DelPopolo:2013qba, DelPopolo:2016emo, Liu:2017drf, Rodrigues:2017vto}. Therefore, phenomena associated to dark matter have a chance of serving as a window towards new physics.

Among the observational correlations whose explanation may be simpler within alternative theories of gravity, the Radial Acceleration Relation\cite{McGaugh:2016leg} (RAR) stands out\cite{Lelli:2017vgz}. It is a sharp correlation between the observed acceleration along the galactic radius $a(R)$ (found from the redshift of atomic or ionised hydrogen) and the Newtonian acceleration of baryonic matter $a_\mscript{N}(R)$ (found from the  mass profiles that are derived from the observed surface brightness of the different galaxy components). Both these accelerations can be derived from observational data independently of any hypothesis on dark matter. From the $\Lambda$CDM perspective, recent results using complex large-scale hydrodynamical simulations\cite{Ludlow:2016qzh, Navarro:2016bfs} indicate that $\Lambda$CDM may reproduce such correlation. This work supports this line of investigation, but we stress that, within the standard model, the RAR  is far from obvious.
The RAR is also known as ``Mass Discrepancy-Acceleration Relation'' (MDAR).

The Modified Newtonian Dynamics (MOND)\cite{1983ApJ...270..371M, Sanders:2002pf, Famaey:2011kh, Milgrom:2015ema} -- although cannot explain so much of the observed universe as the $\Lambda$CDM model does -- correctly predicts the existence of certain features that are not so clear within the standard model, such as the Tully-Fisher relation\cite{1998ApJ...508..132D, Gentile:2009bw, McGaugh:2016leg}.
An essential assumption  of this approach is the existence of a fundamental acceleration scale, given by a constant $a_0$, such that the gravitational force decays with distance slower than Newtonian for accelerations about or smaller than $a_0$.
More precisely, MOND was developed based on the following correction to the Newtonian acceleration in the context of disk galaxies\cite{1983ApJ...270..371M, Sanders:2002pf, Famaey:2011kh}:
\begin{equation} \label{eq:mond}
	\boldsymbol{a}_\mscript{N} = \mu\left(\frac{a}{a_0} \right ) \boldsymbol{a} \,,
\end{equation}
where $\boldsymbol{a}_\mscript{N}$ is the Newtonian acceleration, $\boldsymbol{a}$ is the one expected to be the physical acceleration and $\mu(x)$ is called the interpolating function. This function is such that for large accelerations ($a \gg a_0$) one would find Newtonian gravity ($a=a_\mscript{N} $), while for small ones one gets $a^2 \propto a_\mscript{N}$.
There is not much room to change these limiting cases since $a = a_\mscript{N}$ is required by Solar System dynamics and $a^2 \propto a_\mscript{N}$ is essentially demanded by the Tully-Fisher relation in the absence of dark matter\cite{Tully:1977fu, McGaugh:2000sr}.
The function $\mu(x)$ controls the smoothness of the transition between Newtonian gravity and the regime of very small accelerations. A too sharp transition between these regimes will lead to issues on galaxy dynamics, since  observational data indicates that the transition is smooth along the galaxy radius. While a too smooth transition is more prone to be observed via local tests of gravity\cite{Famaey:2011kh}. The Simple Interpolating Function faces difficulties with Solar System constraints\cite{Hees:2015bna}, while the Standard Interpolating Function  has no issues with the Solar System, but performs worse with galaxies\cite{Famaey:2005fd, Gentile:2010xt}. These two interpolating functions  are the most used ones.
The RAR suggests another interpolating function, which comes directly from its results and we call here RAR-inspired Interpolating Function. The latter is an intermediate case for the range of Newtonian accelerations probed by the rotation curves of galaxies.  In this work, we consider these three interpolating functions (see Methods for their  expressions).
As we will discuss, our conclusions are robust as far as these three interpolating functions are concerned, and likely as far as any other interpolating function that is compatible with the RAR.

The standard procedure to measure $a_0$ from galaxy rotation curves is as follows\cite{Gentile:2010xt, Famaey:2011kh}: first one performs fits on several galaxies considering $a_0$ as a free parameter, then one estimates the true $a_0$ value by taking the median (an estimator more robust than the mean). The quoted error is the error on the median\cite{Gentile:2010xt}, which scales as $1/\sqrt{N}$ where $N$ is the size of the galaxy sample.
The result of this procedure applied to the SPARC sample is presented in the Supplementary Material. Proceeding in this way, one will always find a best $a_0$ value for each interpolating function, but the existence of a fundamental universal $a_0$ is not  tested, it is assumed from the beginning.

MOND is commonly criticized at larger than galactic scales, where its predictions fail to hold\cite{Dodelson:2011qv}, or some form of additional matter seems to be necessary\cite{Sanders:2002ue, Angus:2008qz}. There is hope that a more complete theory, that becomes similar to MOND at galactic scales, may have better results at all scales\cite{Milgrom:2009gv, Babichev:2011kq, Verlinde:2016toy, Smolin:2017kkb}.    For the Milky Way alone, assuming the Standard Interpolating Function and a large class of viable baryonic models, it was shown that the inferred $a_0$ in the Milky Way is incompatible with the one inferred from external galaxies at more than\cite{Iocco:2015iia} $5\sigma$.  This alone is an interesting result, but it has relevant limitations since this conclusion does not hold for the Simple Interpolating Function and it depends on the analysis of a single galaxy (a special galaxy, but one nonetheless). There are other works that criticize MOND by presenting data that suggest a non constant $a_0$ value, but these criticisms were not based on the Bayesian analysis of a large sample of galaxies\cite{Randriamampandry:2014eoa} (see also references therein).

Here, we test whether galaxy data is consistent with a fundamental acceleration scale. If it is not, then the RAR is an emergent phenomenon that appears when diverse galaxies are stacked together.  In this case, MOND, as any other theory that behaves like it at galactic scales, is inconsistent with the data. 

\begin{figure} 
\begin{centering}
\includegraphics[width=\columnwidth]{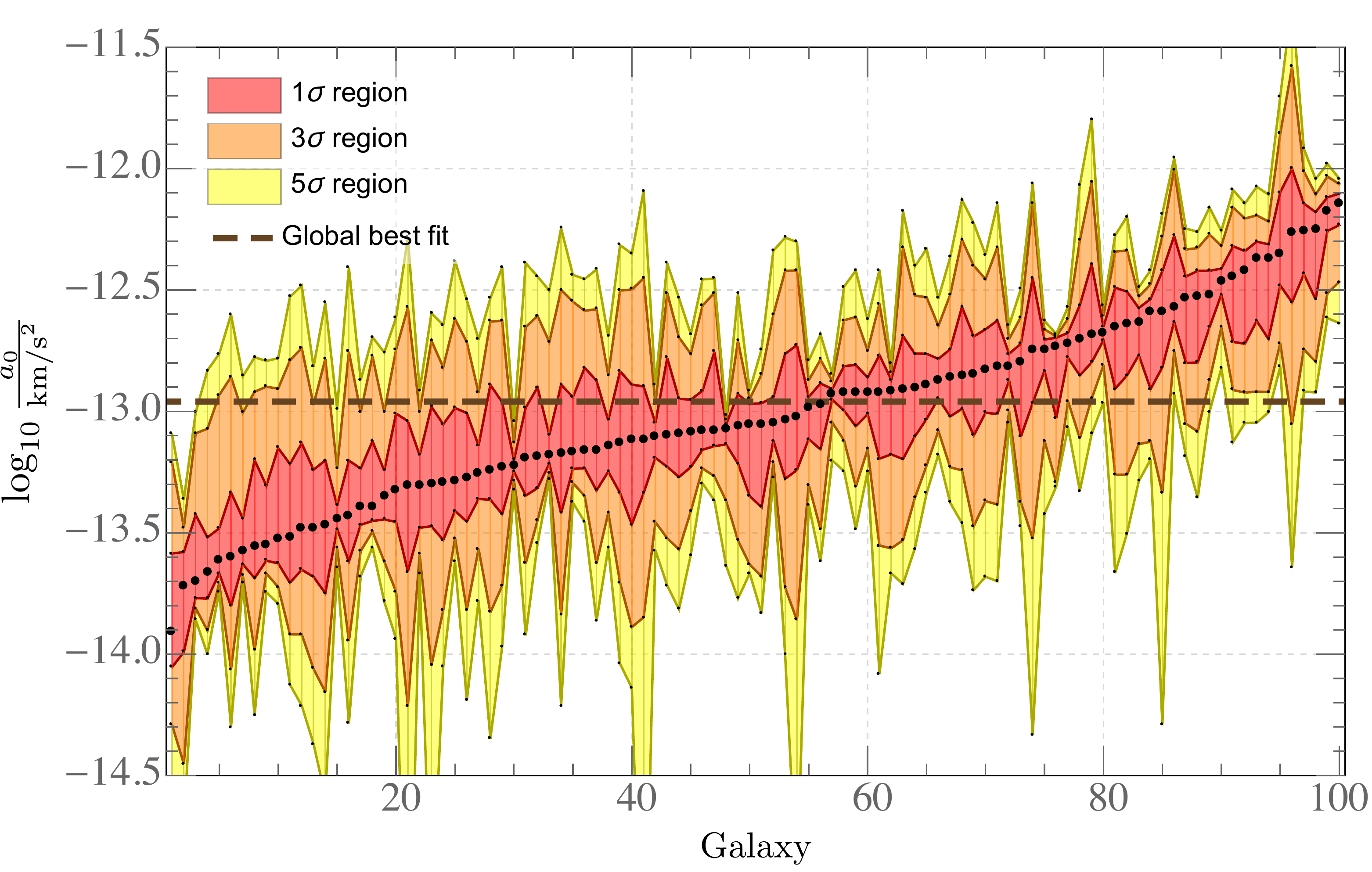}
\caption{
{\bf Posterior probability distributions of $a_{0}$ for the galaxies of the SPARC database.}
Each posterior has been obtained after marginalizing over the allowed ranges of stellar mass-to-light ratios and galaxy distance, and are displayed showing the maximum of the posterior (black dots) together with the 1, 3, and 5$\sigma$ credible intervals.
To enhance the clarity of the plot, the galaxies are sorted according to their posterior maximum.
Only the 100 galaxies that passed the quality criteria are displayed.
The global best-fit value of $a_{0}$ (see Methods) is shown with a dashed line: it is evident that many galaxies are not quite compatible with the global best fit, see Table~\ref{tab:bayesiansigma} for a numerical summary.
The result above is for the  RAR-inspired Interpolating Function, the other interpolating functions lead to similar results and can be found in the Supplementary Material.}
\label{fig:MONDRARBands}
\end{centering}
\end{figure}

A straightforward  way to assess the existence of a fundamental $a_0$ is to perform a full Bayesian analysis and obtain the posterior probability distribution of $a_{0}$ for each galaxy in the sample. One can then test if the various posteriors are compatible with each other.
Bayesian inference has the great advantage that parameters are never fixed to some central or best-fit values and the posterior on $a_{0}$ is obtained by marginalizing over the other parameters. As such, we do not bias the result in any way: any possible degeneracy, expected or unexpected, linear or not, will be correctly taken into account.
Furthermore, we do not linearize the likelihood with respect to the parameters and we obtain the confidence intervals by integrating the posterior distribution without adopting any Gaussian approximations. 
Further details of the analysis (including a recap of Bayesian inference) can be found in the Supplementary Material.

In order to obtain the (unmarginalized) posterior on $a_{0}$, stellar mass-to-light ratios $\Upsilon_{\star d}$ and $\Upsilon_{\star b}$ (for the disk and the bulge, respectively) and galaxy distance $D$, we adopt the following flat priors: $a_{0}>0$, $D$  constrained to lie within $\pm 20\%$ of the observed distance, and $\Upsilon_{\star \mscript{d}}$ and $\Upsilon_{\star \mscript{b}}$ constrained to lie within a factor $2$ of their expected values.
Changes on distance and mass-to-light ratios in general have significant impact on MOND results, since its dark matter-like effects come directly from the usual matter distribution. Hence, to properly evaluate MOND, it is important to consider a wide-enough prior on these parameters. Our choice is of the same order, or larger, than the expected ranges for the SPARC data\cite{McGaugh:2016leg}.
It is important to stress that stellar mass-to-light ratios and galaxy distance are treated here as nuisance parameters that model the impact of possible systematics.
More precisely, as shown in the Supplementary Material, stellar mass-to-light ratios and galaxy distance are correlated with each other and with $a_{0}$. Therefore, by marginalizing over the these nuisance parameters we are effectively increasing the error bars in a manner that is proportional to the widths of the priors we adopt.

Our main results are shown in Figure~\ref{fig:MONDRARBands} and Table \ref{tab:bayesiansigma}. Figure \ref{fig:MONDRARBands} shows the  posterior distributions on $a_{0}$, including their maxima and the 1, 3, and 5$\sigma$ confidence ranges (credible intervals). These results use no Gaussian or linear approximations. All the 175 SPARC galaxies\cite{2016AJ....152..157L} were analysed, but only 100 of them passed the quality cuts (see Methods) and were considered for the analyses of Figure~\ref{fig:MONDRARBands} and Table \ref{tab:bayesiansigma}.
As it is evident, there is no value of $a_0$ that is inside all the $5\sigma$ intervals. Figure~\ref{fig:MONDRARBands} also shows the global best fit of $a_0$ (see Methods). It implies that 13 galaxies, among the selected 100 galaxies,  are incompatible with the  global best fit of $a_0$ by more than $5\sigma$.
This issue is further detailed in Table~\ref{tab:bayesiansigma}.
The existence of a fundamental universal acceleration scale can be discarded with a confidence of 10$\sigma$ or more, as detailed in Methods.

The results above are quite robust. As described in Methods, our conclusions remain unchanged after applying strict quality cuts (we removed 55\% of the sample with the most strict quality criteria, see the Supplementary Material for the plot). Only if there were a strong and systematic underestimation of the errors then the results above could be somehow questioned. However, such a large increase would raise several other issues on rotation curve analyses. Among others, the RAR analysis itself indicates that the average expected errors are about or larger than the observed dispersion\cite{McGaugh:2016leg}, hence increasing the errors would not improve the RAR significance. Moreover, in order to address possible systematic issues that correlate with galaxy luminosity or surface brightness, we also evaluate correlations of our Bayesian results with these parameters, as detailed in the Methods section and in the Supplementary Material. We find no significative correlation capable of changing our results.

\begin{figure}
	\centering
	\begin{subfigure}{0.50\textwidth}
		\centering
		\includegraphics[width=9cm, trim ={0 0.8cm 0 0}, clip]{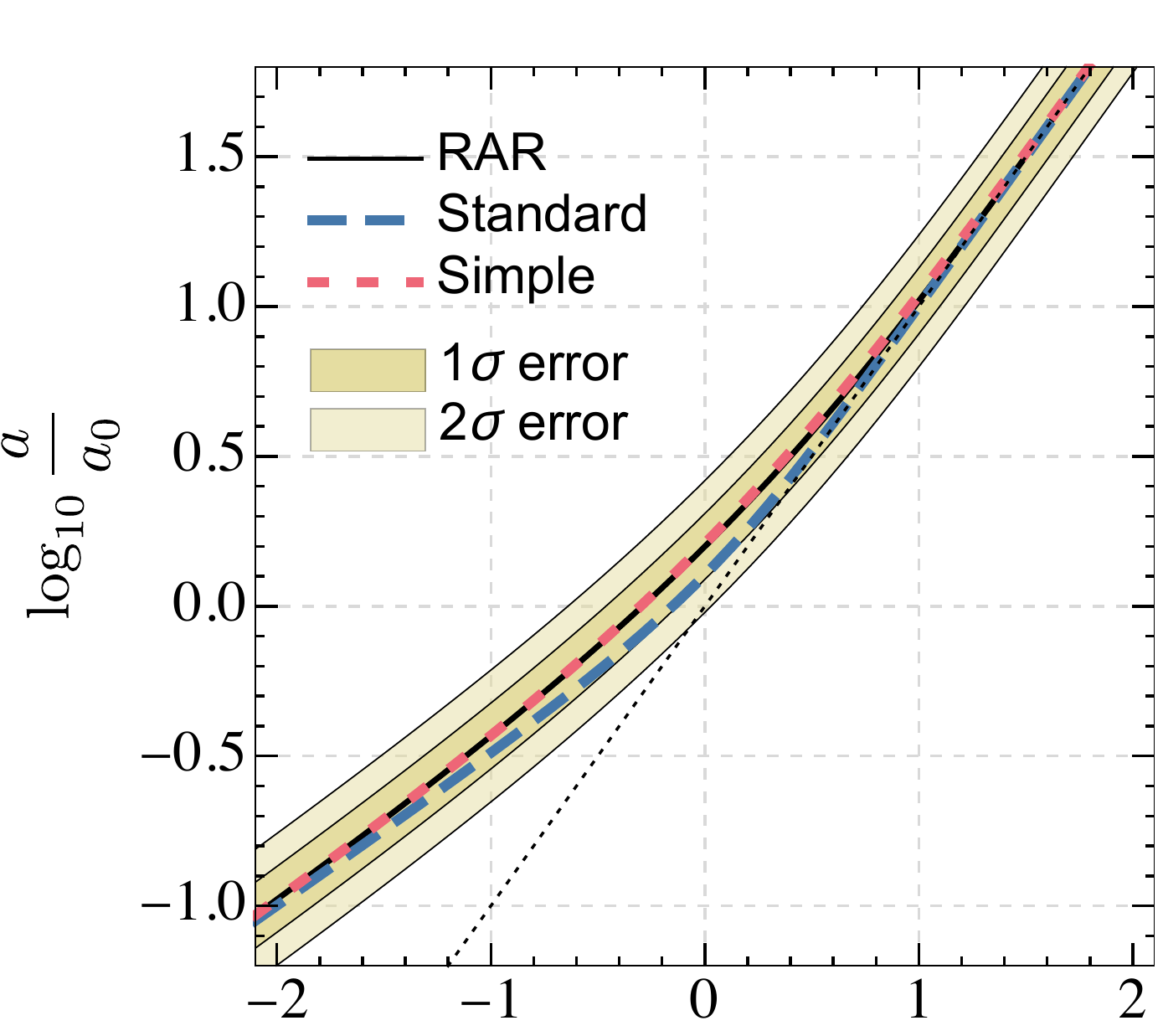}
	\end{subfigure}\\
	\centering
	\begin{subfigure}{0.50\textwidth}
		\centering
		\includegraphics[width=9cm,trim ={0 0 0 0.4cm}, clip]{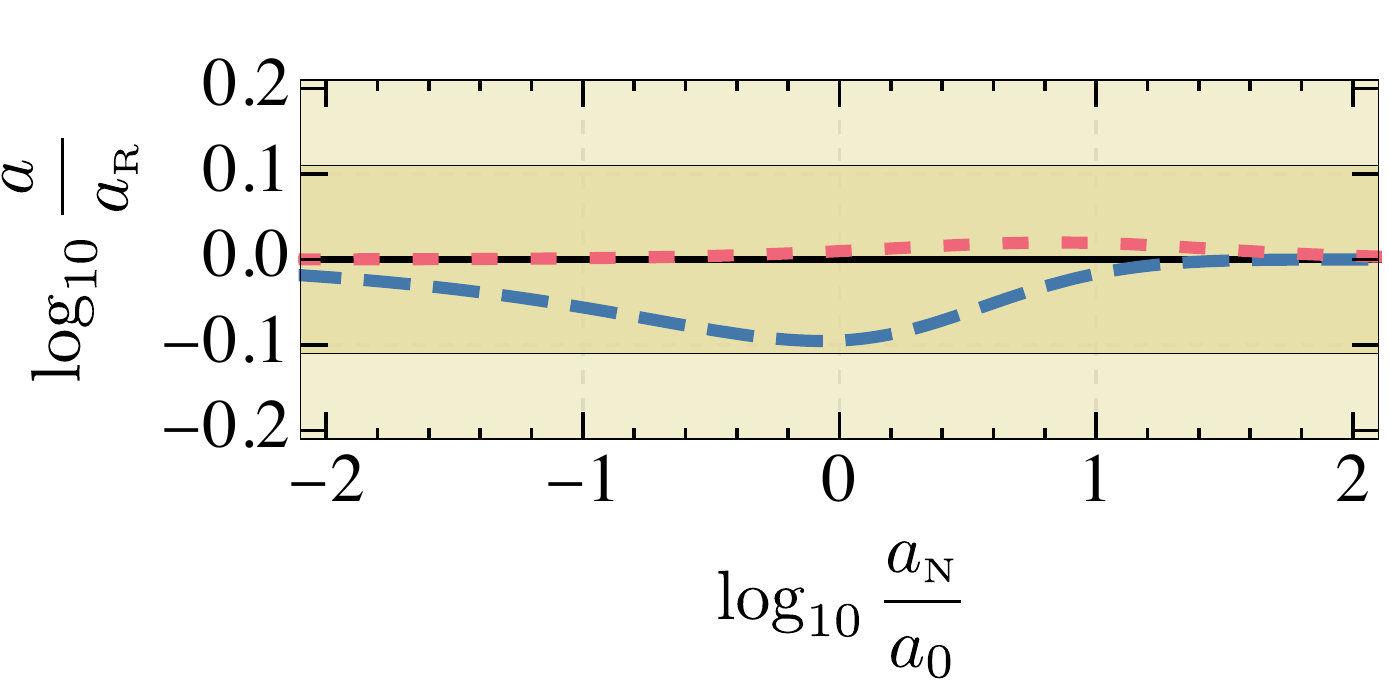}
	\end{subfigure}
	\caption{{\bf The Radial Acceleration Relation (RAR), in units of $a_0$, compared to  the Standard and Simple interpolating functions.}  The absolute value of the observational acceleration is denoted by $a$,  $a_\mscript{N}$ is the Newtonian acceleration inferred from the baryonic matter distribution alone (no dark matter), and $a_\mscript{R}$ is the average acceleration inferred from the RAR. The average 1 and $2\sigma$ dispersions about the RAR  data are  shown as yellowish regions. Using  $a_0 \sim 10^{-13} \mbox{ km/s}^2$, the  data presented by McGaugh et al\cite{McGaugh:2016leg} ranges from about -2 to +2 in the horizontal axis of this plot. The dotted line corresponds to $a = a_\mscript{N}$.} 
	\label{fig:mus}
\end{figure}

The results shown in Figure~\ref{fig:MONDRARBands} were found using the RAR-inspired Interpolating Function, but the Simple and the Standard interpolating functions yield similar results, as summarized in Table \ref{tab:bayesiansigma} and explicitly shown in the Supplementary Material.
This is due to the fact that, as shown in Figure \ref{fig:mus}, the three interpolating functions are inside the $1\sigma$ dispersion of the RAR.
Other possible interpolating functions cannot differ too much from the three interpolating functions we consider, otherwise they will not satisfy the RAR. In the Supplementary Material we also perform  the same analysis for some galaxies of the THINGS sample\cite{2008AJ....136.2648D}. Albeit with less galaxies, the analysis leads to the same conclusion: there is no common acceleration $a_0$ for all the galaxies.

\begin{table}
\caption{\small {\bf Global best-fit $a_0$ values and inconsistencies with a fundamental acceleration scale.} The $\log_{10} a_0$ column lists  the corresponding global best fits (in km/s$^2$) for each interpolating function.  The last three columns show the percentages of galaxies whose  credible intervals at $1, 3 \mbox{ or } 5 \sigma$ on $\log_{10}a_0$ do not intercept the global best fit.
Only the 100 galaxies that passed the quality criteria are considered. Reference values  are given in parenthesis: clearly there are many more incompatible galaxies than expected.}
\begin{center}
\begin{tabular}{l|cccc}
\hline
\hline
  Interpolating &   \multicolumn{4}{c}{Bayesian approach} \\
function &   & \multicolumn{3}{c}{\small incompatible galaxies (expected)}  \\ \cline{3-5} 
   & $\log_{10} a_0  $ & $1\sigma$ ($32\%$)&$3\sigma$ ($0.27\%$)&$5\sigma$ ($5.7\times 10^{-7}$) \\
  \hline
  Standard &-12.902 & 67\% & 30\%& 13\% \\
  Simple & -12.918& 66\% & 28\%&11\% \\
  RAR-inspired & -12.959& 64\%& 28\% & 13\%\\
\hline
\hline
\end{tabular}
\end{center}
\label{tab:bayesiansigma}
\end{table}

Table~\ref{tab:bayesiansigma} summarizes the visual results of Figure~\ref{fig:MONDRARBands} by presenting the percentage of galaxies that are incompatible with the global best fit for $a_0$  at different $\sigma$ levels. Reference values are given in parenthesis: there are clearly many more incompatible galaxies than expected.

Concluding, the RAR indicates the existence of an acceleration scale that emerges from the analysis of several galaxies, when their data is stacked together. On the other hand, from the combination of the individual galaxy analyses, we show, using Bayesian inference, that this emergent acceleration cannot be considered a fundamental acceleration, since the individual results coming from each galaxy are not compatible with a single acceleration.
Our findings imply that MOND, although useful for disclosing certain general properties of galaxies, cannot be a fundamental theory, even at the level of galaxies. Consequently, any proposed theory that extends MOND while preserving  its  dynamics at the level of galaxies cannot be correct (it is rejected at more than 10$\sigma$). 
Due to its emergent nature, the RAR may be the result of a complex interplay between dark matter and baryons. Particular realizations of this picture are currently being investigated\cite{Ludlow:2016qzh, Navarro:2016bfs, Famaey:2017xou}.

\section*{Methods}

\subsection{Data.}
The main data that we use come from the 175 galaxies of the  SPARC sample\cite{2016AJ....152..157L}.
In order to probe possible sample systematics, we also consider 18 galaxies from another high quality sample, the THINGS sample\cite{2008AJ....136.2648D}. The results are qualitatively the same (further details in the Supplementary Material).

\subsection{Data quality and assumptions in the data.} The data samples that we use are commonly cited and used as the best samples for disc galaxy analyses. Nonetheless, we clarify that they are not raw observational data, since they depend,  in particular, on the galaxy surface brightness decomposition into gas, stellar disc and stellar bulge, interpretations on the inclinations, distances, assumptions on axial symmetry and on choices on data binning. When possible we reduce the dependence of our results on such hypotheses; this is done by using nuisance parameters with flat priors, considering different sets of quality cuts and considering two different samples of data (SPARC and THINGS) which use different approaches. Finally, uncontrolled systematics are highly unlikely to appreciably change the strong significance of our results.

\subsection{New numerical packages.}
To analise the data of all the 175 galaxies from SPARC many times with different hypothesis, we developed  two new packages to automatize these procedures. Both packages can be used for other samples, besides SPARC, but are optimized for the latter. They are based on the Wolfram Language and are named MAGMA (for Mathematica Automatized Galaxy Mass Analysis) and mBayes (Bayesian analysis with Mathematica). The MAGMA package has three main functions, and all of them automatically analyse all the galaxies with a single human interaction. The functions have the following purposes: $i$) one for finding the best fit of a model, by minimizing the $\chi^2$ with given constraints, $ii$) another function for generating plots for the individual galaxies with the derived fitted parameters, and $iii$) a last function for generating a single table with relevant data from all the fits (including results from mBayes). The best fit is searched using a differential evolution algorithm many times, with different numerical options in parallel, in a two-step procedure, such that the second step refines the first step results. The main purpose of the mBayes package is to perform the Bayesian analysis and find the confidence contours for all the galaxies. It is the most demanding computational part, and it also uses parallel computing. It imports the $\chi^2$ functions generated by MAGMA
together with the values of the parameters' best fits. The best fits are only used as an initial input for the exploration of the parameter space. The mBayes package generates triangular plots for each galaxy (see Supplementary Materials for two examples) and a table with the numerical values of the 1-to-5$\sigma$ confidence intervals (which is what is used to generate Figure~\ref{fig:MONDRARBands} and Table \ref{tab:bayesiansigma}).

\subsection{Newtonian acceleration and circular velocity.}
The Newtonian acceleration is  by definition $\boldsymbol{a_\mscript{N}} = - \boldsymbol{\nabla} \Phi_\mscript{N}$. The Newtonian potential is defined from the Poisson equation $\nabla^2 \Phi_\mscript{N} = 4 \pi G \rho$, where $G$ is the gravitational constant and $\rho$ the matter density profile. Galaxy matter densities that can be inferred from observations are here decomposed, in accordance with the SPARC and THINGS conventions, into stellar and gaseous components. The stellar component, when necessary, is further decomposed into bulge and disk.
Consequently, the total mass profile of a galaxy is decomposed into bulge, disk and gas according to $\rho = \rho_{\star \mscript{b}} + \rho_{\star \mscript{d}} + \rho_{\mscript{gas}}$. These are the baryonic contributions,  here we consider no dark matter. Due to the linearity of the Poisson equation, for each of these componentes one can derive a corresponding acceleration contribution. Assuming that a disk galaxy is axisymmetric and rotationally supported, the total acceleration experienced by a small radial interval in a galaxy  must be oriented towards the galaxy center. Since matter is not spherically  distributed in a disk galaxy, individual contributions to the total Newtonian acceleration may be oriented to the opposite direction.
These negative contributions are relevant for some galaxies, and are considered in our codes.
If the acceleration of a given galaxy component is towards the galaxy center, then $a \equiv |\boldsymbol{a}|= V^2/R$ for that component,  where $R$ is the cylindrical radial coordinate. In the context of galaxy rotation curves, it is common to adopt the following convention valid for any direction of the acceleration: $V = - a_R  \sqrt{ R/ |a_R |}$, where $a_R$ is the radial component of the acceleration. Hence, galaxy components that reduce the total acceleration at a given radius are represented by a negative velocity at that radius.
See the Supplementary Material for examples of galaxy rotation curves.

Both the SPARC and the THINGS databases provide the circular velocities of each baryonic component, and with the sign convention above. The square of the total Newtonian circular velocity is written as
\begin{equation} \label{v2}
	V^2_\mscript{N} \equiv \Upsilon_{\star \mscript{b}}|V_{\star \mscript{b}}| V_{\star \mscript{b}}+ \Upsilon_{\star \mscript{d}}|V_{\star \mscript{d}}| V_{\star \mscript{d}} + |V_{ \mscript{gas}}| V_{ \mscript{gas}}.
\end{equation}
Due to the uncertainties on the stellar mass-to-light ratios, the dimensionless mass-to-light ratio constants are inserted in the expression above. The provided data on the circular velocity components are for $\Upsilon_\star = 1$ (this value is adopted for the data presentation in order to easy the conversion to any other mass-to-light ratios).

\subsection{Models: MOND with three different interpolating functions.}
All the analyses performed here are done considering MOND, but with three different interpolating functions $\mu(x)$, see equation~(\ref{eq:mond}). For the model analysis, it is useful to invert equation~(\ref{eq:mond}) and express the acceleration $\boldsymbol{a}$ as a function of $\boldsymbol{a_\mscript{N}}$. It is in this form that the interpolating functions are implemented in the code. For the Standard, Simple and RAR-inspired interpolating functions, the  accelerations read:

\begin{eqnarray} \label{eq:interpol}
\boldsymbol{a}_\mscript{std} = \frac{\boldsymbol{a_{\mscript{N}}}}{\sqrt 2}\sqrt{1 + \sqrt{4 \frac{a_0^2}{a_{\mscript{N}}^2} +1} } \,, \;\;\;
\boldsymbol{a}_\mscript{smp} = \frac{\boldsymbol{a_{\mscript{N}}}}{2}\left(1 + \sqrt{4 \frac{a_0}{a_{\mscript{N}}} +1} \, \right) \,,\;\;\;
\boldsymbol{a}_\mscript{rar} = \frac{\boldsymbol{a_{\mscript{N}}}}{1 - e^{-\sqrt{a_\mtiny{N}/a_0}}} \,.
\end{eqnarray}

\subsection{Fitted parameters and $\chi^2$ function.}
For all galaxies we consider a tolerance of $20\%$ for the galaxy distance and a tolerance of a factor two on the mass-to-light ratios (both for the disk and the bulge, when present). For the SPARC sample, the expected values for $\Upsilon_{\star d}$ and $\Upsilon_{\star b}$ are respectively 0.5  and 0.7 $M_\odot/L_\odot$ in the 3.6 $\mu m$ band (these are fixed for all the galaxies).
These parameters, together with the constant $a_0$, compose the fitted parameters. The best fit relative to a given galaxy is defined as the value that maximizes the likelihood function $\mathcal{L} \propto e^{-\chi^{2}/2}$ or, equivalently, minimizes the $\chi^{2}$ function,
\begin{equation}
\chi^{2}(a_{0},\Upsilon_{\star \mscript{d}}, \Upsilon_{\star \mscript{b}},\delta )= \sum_{i=1}^{N_{g}} \frac{(V(R_{i},a_{0},\Upsilon_{\star \mscript{d}}, \Upsilon_{\star \mscript{b}},\delta )-V_{\mscript{Obs},i})^{2}}{\sigma_{i}^{2}} \,,
\end{equation}
where $N_{g}$ is the number of data points, $R_{i}$ is the radius at which the velocity $V_{\mscript{Obs},i}$ was measured with the error $\sigma_{i}$, and $V$ is the theoretical velocity from MOND (with a given interpolating function). The relation between the theoretical physical acceleration $a$ according to MOND and the velocity $V$ is  $V^2/R = a$.

Details of the Bayesian analysis are in the Supplementary Material.

\subsection{Quality cuts.}\label{sec:datasel}

We perform the following four quality cuts; see the Supplementary Material for the names of the galaxies that were excluded by each quality cut.
First, we remove galaxies that have a very high minimum $\chi^{2}$: in these cases MOND is already a bad model -- perhaps because of specific observational or dynamical issues -- and it does not make sense to include them in the present analysis as the null hypothesis assumes that MOND does work for individual galaxies.
A consistent way to achieve the latter is to set a fixed $p$-value below which we reject the (null) hypothesis that MOND provides a good fit. Specifically, in the main analysis we decide not to analyze galaxies for which MOND is rejected at the 5$\sigma$ confidence level, that is, we exclude galaxies for which:
\begin{equation}
1 - p\text{-value} \equiv F_{k}(\chi^{2}) \ge \erf \frac{n=5}{\sqrt{2}} \,, 
\end{equation}
where $F_{k}(\chi^{2})$ is the cumulative $\chi^2$-distribution with $k$ degrees of freedom, $\chi^{2}$ is the observed value and $\erf$ is the error function.
In this case it is $k=N_{g}-M$ where $N_{g}$ is the number of data points and $M$ the number of fitted parameters, which is 3 or 4 depending on the presence of the bulge.
The reference $p$-value above is $5.7\times 10^{-7}$.
This selection eliminates 37 galaxies.
In order to increase the robustness of our analysis, we also consider the stricter selection criterion of not analyzing galaxies for which MOND is rejected at the 3$\sigma$ confidence level (reference $p$-value of $2.7\times 10^{-3}$).
This eliminates 62 galaxies.

Secondly, we do not include in the analysis of Figure~\ref{fig:MONDRARBands} and Table \ref{tab:bayesiansigma} galaxies whose posterior on $a_{0}$ does not go to zero for $\log a_{0} \rightarrow - \infty$ so that the $5\sigma$ regions are ill defined.
These galaxies accept very low values of $a_0$ and  would not improve the chances of a fundamental $a_0$ value. 
This second selection eliminates  27 galaxies.

Thirdly, we do not include the 12 SPARC galaxies that are marked with the quality flag ${\cal Q} = 3$ (poor quality, as defined by SPARC).
Finally, other 10 SPARC galaxies whose inclination is less than $30^{\circ}$ are  not considered. 
These last two criteria  give the ``RAR flag'' ($\cal R$), which is 1 for the 153 SPARC galaxies that were considered when evaluating the RAR\cite{McGaugh:2016leg}, and zero otherwise.
Due to the elimination of the (close to) face-on galaxies, we do not consider galaxy inclination errors since they have minor impact for\cite{2016AJ....152..157L} $i \geq 30^\circ$. The uncertainties on stellar mass-to-light ratios and galaxy distance are relevant for the dynamical analysis, and these are taken into consideration through the flat priors on $\Upsilon_\star$ and $D$, as described above.

After these four quality cuts, we are left with 100 galaxies.
While 79 galaxies are left in the case of the stricter 3$\sigma$ criterion on the minimum $\chi^{2}$.

\subsection{Global best fit.}

For each galaxy we compute the mean and the variance of the posterior according to:
\begin{align}
\bar a_{0,j} &= \int \d a_{0} \, a_{0} \, p_{j}(a_{0}) \,,\\
\sigma_{a,j}^{2} &= \int \d a_{0} \, (a_{0}-\bar a_{0,j})^{2} \, p_{j}(a_{0}) \,,
\end{align}
where $p_{j}(a_{0})$ is the (marginalized) posterior on $a_{0}$ for the galaxy $j$, with $1\le j \le N$ where $N$ is the number of galaxies that passed the quality cuts discussed above.
Then, in order to find the global best-fit value of $a_{0}$ we introduce the following $\chi^{2}$ statistic:
\begin{equation} \label{chi2a}
\chi^{2}(a_{0}) = \sum_{j=1}^{N} \frac{(\bar a_{0,j}-a_{0})^{2}}{\sigma_{a,j}^{2}} \,,
\end{equation}
which follows a $\chi^{2}$-distribution with $k=N-1$ degrees of freedom, under the approximation that $\bar a_{0,j}$ is distributed according to a Gaussian distribution with mean $\bar a_{0,j}$ and variance $\sigma_{a,j}^{2}$.
The minimization of $\chi^{2}(a_{0})$ yields the ``global best fit'' $a_{0}^{\rm bf}$. 

\subsection{Confidence level in rejecting MOND.}\label{sec:nomond}

As the degrees of freedom $k$ of the $\chi^{2}$ defined in Equation~\eqref{chi2a} is high ($k\sim 100$), we can approximate the $\chi^{2}$-distribution with a Gaussian with mean $k$ and standard deviation $\sqrt{2 k}$.
We can then estimate the number of $\sigma$ at which MOND is rejected according to:
\begin{equation}
n_{\sigma} = \frac{\chi^{2}(a_{0}^{\rm bf})-k}{\sqrt{2 k}} \,.
\end{equation}
Using any of the three interpolating functions for the SPARC sample we find that $n_{\sigma} \sim 60$ (main analysis, 5$\sigma$ criterion for the exclusion of galaxies with high $\chi^{2}$) and $n_{\sigma} \sim 40$ (3$\sigma$ criterion).

\subsection{Correlations with galactic  parameters and observational data robustness.}
Observational data on galaxy rotation curves may in principle be more trustworthy for certain values of the galaxy parameters  than for others (e.g., some systematic issues with the data may correlate with luminosity, surface brightness, asymptotic circular velocity, \dots). The quality cuts previously described do not explicitly address issues of this type. In particular, in high luminosity galaxies it is more frequent to find local features in the rotation curves,  and these  may be either interpreted as necessary features to be reproduced by galaxy models (like the Renzo rule), or as features that appear due to poor modeling of the baryonic matter\cite{2008AJ....136.2648D, 2016AJ....152..157L} (e.g., a significative violation of axial symmetry in the observational data, or an error in addressing a variation of the inclination). In the latter case, a local disagreement between the model and the  rotation curve data would not be a problem.

In the Supplementary Material we present our analysis of this topic in more detail. We find at most weak correlations that have no significant impact on our results. In particular, by imposing  additional quality cuts, such that galaxies of sufficiently high luminosity are not considered, the tension between the data and the existence of a fundamental acceleration scale can be reduced, but the reduction does not change our main results. Even with these additional quality cuts, a fundamental (universal) acceleration scale is still discarded  at more than 10$\sigma$.


\begin{addendum}
 
\item
We thank Stacy McGaugh for  clarifications regarding the SPARC sample and comments on a previous version of this paper. This work made use of  SPARC ``Spitzer Photometry \& Accurate Rotation Curves'' and of THINGS ``The HI Nearby Galaxy Survey''. 
DCR and VM thank CNPq and FAPES for partial financial support. ADP was supported by the Chinese Academy of Sciences and by the President's international fellowship initiative, grant no. 2017 VMA0044. ZD thanks the ministry of science, research and technology of Iran for  financial support.

\item[Author contributions]
DCR and ADP proposed the study. DCR developed the MAGMA package, performed the $\chi^2$ minimization analysis and contributed to interpretation and design. VM developed the mBayes package, performed the Bayesian analysis and contributed to interpretation and design. ZD carried out the THINGS sample analysis and raised issues that were essential for the beginning of this project. The first draft was written by DCR and VM, and all the authors contributed to its development.
\item[Competing Interests]
The authors declare that they have no competing financial interests.

\item[Materials \& Correspondence]
Correspondence and requests for materials should be addressed to DCR or VM (emails: davi.rodrigues@cosmo-ufes.org, marra@cosmo-ufes.org).

\end{addendum}

\bibliography{mond}

\end{document}